\documentclass{Interspeech}
\usepackage{comment}
\usepackage{booktabs} 
\usepackage{bm}
\usepackage{graphicx}
\usepackage{listings}    

\lstdefinestyle{mystyle}{
    backgroundcolor=\color{},   
    basicstyle=\ttfamily\color{black}\fontsize{10pt}{10pt}\selectfont, 
    breaklines=true,                   
}

\usepackage[dvipsnames]{xcolor} 
\definecolor{myblue}{RGB}{0, 0, 254}      
\definecolor{mygreen}{RGB}{46, 139, 87}    
\definecolor{myred}{RGB}{192, 25, 50}      

\interspeechcameraready

\title{Leveraging LLM for Stuttering Speech: A Unified Architecture Bridging Recognition and Event Detection}

\author[affiliation={1}]{Shangkun}{Huang}
\author[affiliation={1}]{Jing}{Deng}
\author[affiliation={2,3}]{Jintao}{Kang}
\author[affiliation={1}]{Rong}{Zheng}

\affiliation{}{Beijing Fosafer Information Technology Co., Ltd.}{China}
\affiliation{}{Institute of Forensic Science, Ministry of Public Security}{China}
\affiliation{}{The Institute of Linguistics, Chinese Academy of Social Sciences}{China}
\email{huangshangkun@fosafer.com}

\keywords{Mandarin stuttered speech, ASR, SED, Multi-task Learning, LLM}

\makeatletter
\renewcommand\NoHyper{}
\makeatother

\hypersetup{
    colorlinks=true,
    linkcolor=myblue,
    citecolor=blue,
    urlcolor=blue,
    pdfborder={0 0 0},
}

\begin{document}

\maketitle

\begin{abstract}
    
The performance bottleneck of Automatic Speech Recognition (ASR) in stuttering speech scenarios has limited its applicability in domains such as speech rehabilitation. This paper proposed an LLM-driven ASR-SED multi-task learning framework that jointly optimized the ASR and Stuttering Event Detection (SED) tasks. We proposed a dynamic interaction mechanism where the ASR branch leveraged CTC-generated soft prompts to assist LLM context modeling, while the SED branch output stutter embeddings to enhance LLM comprehension of stuttered speech. We incorporated contrastive learning to strengthen the discriminative power of stuttering acoustic features and applied Focal Loss to mitigate the long-tailed distribution in stuttering event categories. Evaluations on the AS-70 Mandarin stuttering dataset demonstrated that our framework reduced the ASR character error rate (CER) to 5.45\% (-37.71\% relative reduction) and achieved an average SED F1-score of 73.63\% (+46.58\% relative improvement).
\end{abstract}

\begin{figure*}[htb]
    \centering
    \includegraphics[width=0.96\textwidth]{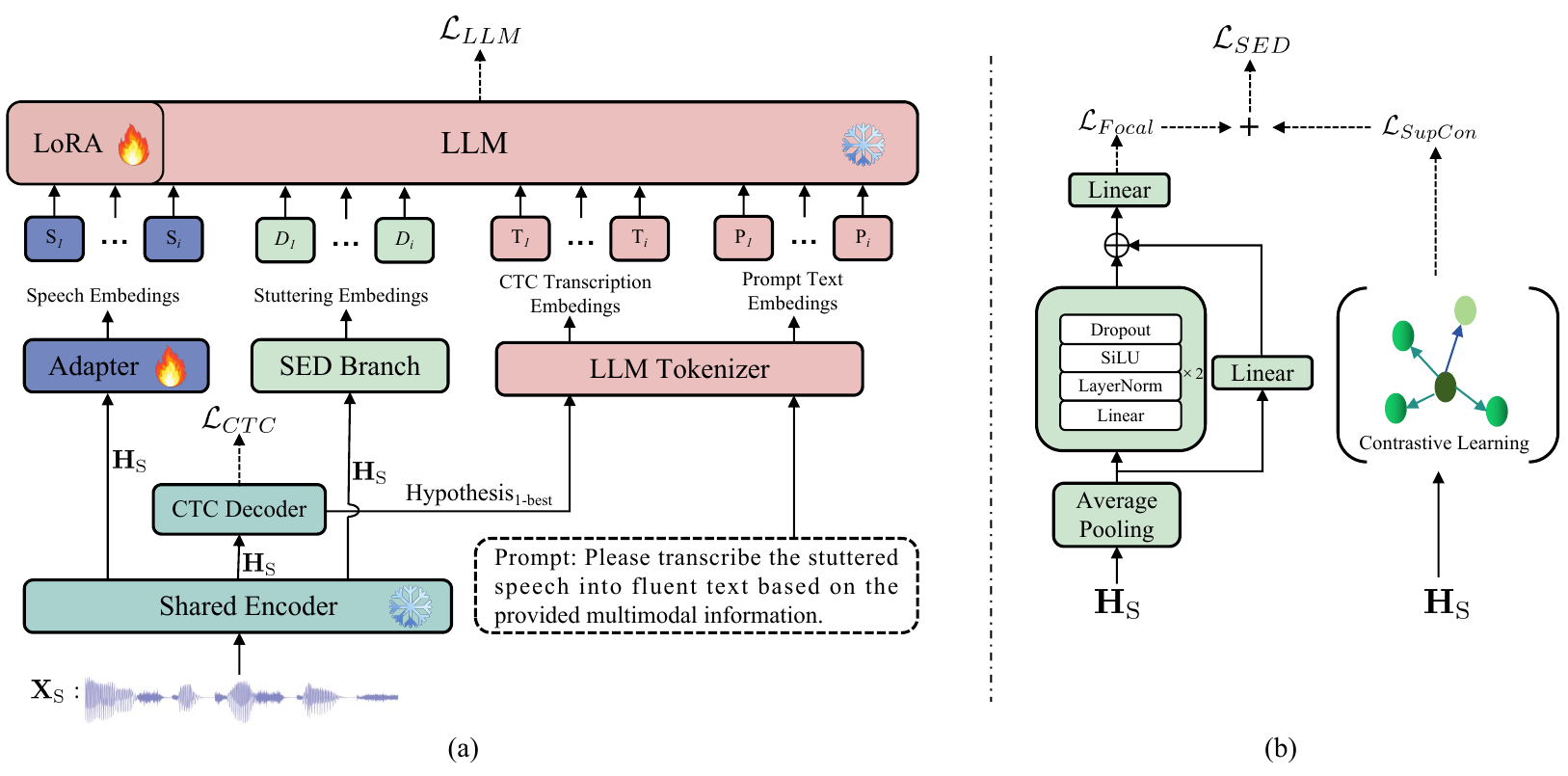} 
    \caption{(a) LLM-driven ASR-SED multi-task learning framework; (b) SED branch architecture}
    \label{fig:dudu1} 
\end{figure*}

\section{Introduction}

Stuttering is a multifaceted neurodevelopmental disorder affecting over 80 million people worldwide\cite{wu2023world}. Common manifestations include word, phrase, or sound repetitions, silent pauses (blocking), elongation of specific syllables or words (prolongation), and interjections. Stuttering impedes daily communication for people who stutter, often leading to stress, shame, and feelings of inferiority. This fear of speaking can lead to self-isolation\cite{coalson2022microaggression}. Early diagnosis and timely intervention are crucial for mitigating the long-term impacts of the disorder, yet many individuals remain undiagnosed until their symptoms become more pronounced. Increased awareness and access to specialized care can help reduce both personal and societal burdens. In mainland China, speech therapy practices are still in their infancy compared to North America and Western Europe, leading to delays in diagnosing and treating stuttering in children.

Despite significant advancements in speech signal processing technology, the performance of ASR in stuttering speech scenarios remains limited, hindering its application in fields such as speech rehabilitation\cite{2-huang2024enhanced,1-huang2024fosafer}. Gong et al. introduced AS-70, the first publicly available dataset of Mandarin stuttered speech, which remains the largest and most comprehensive dataset in this domain \cite{as-70}. The SLT 2024 Mandarin Stuttering Event Detection and ASR Challenge has driven recent progress in stuttering recognition and detection technology\cite{xue2024findings}. Recent studies have advanced research on stuttering speech recognition and event detection. Zayats et al. demonstrated improved ASR performance by categorizing stuttering events\cite{21-zayats2016disfluency}, while Shonibare et al. achieved enhanced results through similar approaches\cite{22-shonibare2022enhancing}. Lea et al. fine-tuned ASR models using data from stutterers and improved voice activity detection techniques to boost performance\cite{23-lea2023user}. Mitra et al. reduced word error rates (WER) by adjusting ASR decoding parameters\cite{24-mitra2021analysis}, while Zhang et al. optimized ASR by generating stuttering speech with text-to-speech techniques \cite{25-zhang2022stutter}. In the area of stuttering event detection, Sheikh et al. proposed a Time Delay Neural Network for this task \cite{sheikh2021stutternet}, while Al et al. employed a two-dimensional atrous convolutional network to capture both spectral and temporal features \cite{al2022stuttering}. Kourkounakis et al. developed a disfluency detection network combining ResNet and BiLSTM architectures\cite{kourkounakis2020fluentnet}. Liu et al. introduced a model integrating the Conformer architecture with LSTM networks, aiming to address challenges in generalization and the limitations of available data in stuttering research\cite{liu2024end}. Contrastive learning, which is widely used in image classification and NLP\cite{cl2-2he2020momentum, cl3-zhang2021cola}, enhances discriminative ability and has potential to advance SED.

Large language models (LLMs), known for their robust language modeling capabilities, have been applied across various domains. Recent research has explored LLM integration with ASR systems, leading to the development of LLM-based ASR frameworks\cite{mu2024mmger,ma2025speech,li2024transcription}. LauraGPT\cite{du2023lauragpt} combines an improved Conformer encoder with Qwen-2B for end-to-end training across multiple speech and audio tasks. SpeechGPT\cite{zhang2023speechgpt} utilizes HuBERT-discretized speech tokens and fine-tunes LLaMA-13B. 
while Qwen-Audio\cite{chu2023qwen-audio} initializes an encoder with the Whisper model and optimizes output loss with a frozen Qwen-7B. SALMONN\cite{tang2023salmonn} integrates Whisper-large and BEATs for speech and audio encoding, employing Q-Former to execute multimodal audio tasks. Studies such as generative error correction \cite{hu2024large} demonstrates that post-processing ASR transcriptions with LLM can also enhance recognition performance.

This paper proposed a multi-task learning framework driven by an LLM that jointly optimized ASR and SED tasks. The primary contributions include: (1) a dynamic interaction mechanism that integrated CTC hypothesis embeddings from the ASR branch with the LLM's context modeling, effectively suppressing stuttering-induced repetitive hallucinations and enhancing the robustness of ASR; (2) the integration of stuttering embeddings from the SED branch to help the LLM better understand stuttered speech and generate fluent transcriptions; (3) a hybrid loss function combining contrastive learning and focal loss to strengthen the discriminative power of stuttering acoustic features and address the long-tail distribution in event categories; and (4) state-of-the-art performance on the AS-70 dataset with a CER of 5.45\% and an average F1-score of 73.63\%, providing an end-to-end unified solution for stuttered speech processing and establishing a technical framework for applications such as speech rehabilitation.

\section{Proposed Method}

\subsection{Framework Overview}

As shown in \autoref{fig:dudu1} (a), the proposed LLM-driven ASR-SED multi-task learning framework comprised two parallel branches: the ASR branch and the SED branch. The framework included six core components: the LLM, the LLM tokenizer, the shared encoder, the Connectionist Temporal Classification (CTC)\cite{ctc-graves2006connectionist} decoder, the adapter module, and the SED branch processing module. Additionally, the SED branch incorporated multi-label supervised contrastive learning and a focal loss function (details in \autoref{fig:dudu1} (b)). The implementation details of each module are described in subsequent sections.

\subsection{ASR Branch}

The ASR branch collaborated with the shared encoder and CTC decoder to dynamically generate the CTC optimal hypothesis. This mechanism suppressed stuttering-induced repetitive hallucinations in the LLM and enhanced the model's ability to process stuttered speech by incorporating text modality information.Given a speech feature matrix $X_S \in \mathbb{R}^{T \times D}$, the shared encoder extracted the acoustic feature representation:

\begin{equation}
H_s = \text{Encoder}(X_S)
\end{equation}

where $T$ denotes the time step and $D$ represents the feature dimension. The CTC decoder performed a greedy search to generate the 1-best hypothesis. The CTC loss was formulated as:
\begin{equation}
\mathcal{L}_{CTC}=-\log\sum_{\pi\in\mathcal{A}(Y_{CTC})}\prod_{t = 1}^{T}P(\pi_t|H_s)
\end{equation}
where $\mathcal{A}(Y_{CTC})$ denotes the set of all valid paths aligned with the target sequence $Y_{CTC}$. During training, only the CTC decoder parameters were updated, while the encoder parameters remained frozen. This dynamic parameter update strategy ensured the temporal adaptability of the optimal hypothesis. The generated text hypothesis was processed by the LLM tokenizer, forming a token sequence with global semantic information:
\begin{equation}
E^{CTC} = \text{Tokenizer}(\text{Hypothesis}_{1\text{-best}})
\end{equation}

which provided the contextual foundation for subsequent LLM modules.

\subsection{SED Branch}
\label{subsec: sed_branch}

To address the acoustic differences between stuttered and fluent speech, the SED branch was designed to generate sentence-level stuttering embeddings that enhanced the LLM's comprehension and fluency in transcription. The processing flow included three stages:

\textbf{Feature Extraction Stage:} The acoustic features $H_s$ output by the shared encoder are averaged through pooling to obtain a context vector $v \in \mathbb{R}^d$.

\textbf{Classifier Architecture:}A two-layer fully connected network with residual connections was implemented. Each layer was followed by LayerNorm, a SiLU activation, and Dropout. In the second layer, the input was directly passed through a residual connection and summed with the output. The final layer mapped features to the event categories.

\begin{equation}
\begin{cases}
h_1 = \text{Dropout}(\text{SiLU}(\text{LayerNorm}(W_1v + b_1)))\\
h_2 = \text{LayerNorm}(h_1 + W_2h_1 + b_2)
\end{cases}
\end{equation}

where $W_*$ denotes the trainable parameter matrix. The output layer mapped the feature vector $h_2$ to the event category space $\mathbb{R}^c$.

\textbf{Hybrid Loss Function}: Focal loss was combined with multi-label supervised contrastive learning.

\begin{equation}
\mathcal{L}_{SED} =  \mathcal{L}_{Focal}(p_t) + \delta \mathcal{L}_{\text{SupCon}} 
\end{equation}

Focal loss addressed class imbalance through the modulation factor  $\gamma$ and balancing factor $\alpha$, downweighting well-classified samples and emphasizing hard examples\cite{focal-ross2017focal}. This strategy enhanced minority class recognition.

\begin{equation}
\mathcal{L}_{\text{Focal}}(p_t) = - \sum_{c=1}^{C} \alpha_c (1 - p_c^t)^\gamma \log(p_c^t)
\end{equation}

Multi-label supervised contrastive learning leveraged label co-occurrence to construct positive pairs: samples sharing at least one event category label were defined as positive pairs. We introduced a nonlinear projection layer (an MLP with ReLU activation) after the shared encoder to map features into a lower-dimensional space, followed by L2 normalization. Unlike traditional contrastive learning relying on data augmentation, our method maximized similarity $z_i\cdot z_j$ for positive pairs $(i, j)$ while minimizing similarity with other samples (including negatives and itself). The contrastive loss was computed using a temperature-scaled cosine similarity matrix to focus on hard negatives:


\begin{equation}
\mathcal{L}_{\text{SupCon}}=-\frac{1}{|\mathcal{P}|}\sum_{(i,j)\in\mathcal{P}}\log\frac{\exp(z_i\cdot z_j/\tau)}{\sum_{k = 1}^{N}\exp(z_i\cdot z_k/\tau)}
\end{equation}

where $z_i$ and $z_j$ are L2-normalized projected features, $\tau$ is the temperature coefficient, and $\mathcal{P}$ denotes all positive pairs.

\subsection{LLM Integration: Multi-Modal Embeddings}


For the speech embedding, inspired by prior work\cite{ma2025speech}, we employed a projection layer to transform the speech feature matrix $H_S$ into an embedding $E^S$ aligned with the LLM's input dimension. To preserve the temporal resolution and feature density of speech signals, we did not apply downsampling. The projection layer comprised a single hidden layer with ReLU activation and a regression layer.

The LLM module integrated five input embeddings: (1) speech embedding $E^S$, (2) stuttering embedding $E^{SED}$, (3) CTC hypothesis embedding $E^{CTC}$, (4) transcribed text embedding $E^{T}$, and (5) prompt embedding $E^{P}$, which is expressed as:

\begin{equation}
\begin{cases}
E^S = \text{Linear}(\text{ReLU}(\text{Linear}(H_S))) \\
E^{SED} = \text{Linear}(H_{SED}) \\
E^{CTC} = \text{Tokenizer}(\text{Hypothesis}_{1\text{-best}}) \\
E^{T} =  \text{Tokenizer}(X^{Text}) \\
E^{P} =   \text{Tokenizer}(X^{Prompt})\\
\end{cases}
\end{equation}

where $H_{SED}$ denotes the intermediate output of the SED classifier, aligned with the LLM's embedding dimension via a linear layer.

The final LLM input representation is as follows:
\begin{equation}
E = 
\begin{cases}
\text{Template}(E^S, E^{SED}, E^{CTC}, E^{T}, E^{P}) & \text{if training,} \\
\text{Template}(E^S, E^{SED}, E^{CTC}, E^{P}) & \text{if inference,}
\end{cases}
\end{equation}

During end-to-end joint training, the model leveraged the LLM's comprehension capability to fuse frame-level speech representations, globally informed CTC hypotheses, and stuttering embeddings. This multimodal integration enabled precise modeling of stuttering patterns across speech and text modalities. We fine-tuned the Qwen2.5-3B-Instruct model using Low-Rank Adaptation (LoRA)\cite{hu2021lora} with a cross-entropy (CE) loss for downstream task adaptation.


Finally, the total loss consists of the LLM loss $\mathcal{L}_{LLM}$, CTC loss $\mathcal{L}_{CTC}$, and SED loss $\mathcal{L}_{SED}$, expressed as:
\begin{equation}
\mathcal{L}_{total} = \mathcal{L}_{LLM} + \beta \mathcal{L}_{CTC} + \mu \mathcal{L}_{SED}
\end{equation}

where $\beta$ and $\mu$ are adjustable hyperparameters.

\section{Experiments}

\setlength{\tabcolsep}{10pt}
\begin{table*}[htb]
    \caption{Comparison of CER (\%) between the method we proposed and existing methods for stuttering speech recognition}
    \centering
    \begin{tabular}{lcccccc}
        \toprule
        \textbf{Model} & \textbf{Mild} & \textbf{Moderate} & \textbf{Severe} & \textbf{Conversation} & \textbf{Command} & \textbf{All} \\
        \midrule
        \multicolumn{7}{l}{\textbf{\textit{Previous works (NN-based ASR)}}} \\
        \midrule
        \textbf{Whisper-large-v2}\cite{as-70} & 5.18 & 13.46 & 18.92 & 10.19 & 5.74 & 8.75 \\
        \textbf{Hubert-large}\cite{as-70} & 6.25 & 9.67 & 7.85 & 9.75 & 2.03 & 7.25 \\
        \textbf{Conformer}\cite{as-70} & 5.20 & 8.00 & 9.25 & 7.96 & 3.32 & 6.24 \\
        \midrule
        \multicolumn{7}{l}{\textbf{\textit{Ours (LLM-driven ASR-SED multi-task learning framework)}}} \\
        \midrule
        \textbf{Speech embedding} & 7.97 & 12.24 & 15.87 & 10.03 & 9.69 & 10.13 \\
       \hspace{1.5em}\textbf{+ CTC hypothesis embedding} & 5.06 & 6.97 & 5.78 & 7.24 & 1.48 & 5.63 \\
            \hspace{3em}\textbf{+ Stuttering embedding} & \textbf{4.96} & \textbf{6.64} & \textbf{5.46} & \textbf{7.06} & \textbf{1.32} & \textbf{5.45} \\
        \bottomrule
    \end{tabular}
    \label{tab:asr}
\end{table*}

\subsection{Experimental Configuration}


For the shared encoder, we employed the SenseVoice-Small\cite{4-an2024funaudiollm} model\footnote{https://huggingface.co/FunAudioLLM/SenseVoiceSmall}, which included a SAN-M-based encoder, task-specific embeddings, and a CTC loss function. This non-autoregressive architecture provided efficient low-latency multilingual speech processing. We selected Qwen2.5-3B-Instruct\footnote{https://huggingface.co/Qwen/Qwen2.5-3B-Instruct}, a model pre-optimized for command-following tasks, as the LLM backbone.

During training, we adopted the following data organization format: ``\lstinline[style=mystyle]{<|im_start|>user<S><D><C><P><|im_end|><|im_start|>assistant<T><|im_end|>|}", where \lstinline[style=mystyle]{<S>} represents the speech embedding, \lstinline[style=mystyle]{<D>} represents the stuttering embedding, \lstinline[style=mystyle]{<C>} represents the hypothesis embedding, \lstinline[style=mystyle]{<P>} represents the prompt embedding, and \lstinline[style=mystyle]{<T>} represents the corresponding transcription text embedding. The markers \lstinline[style=mystyle]{<|im_start|>} and \lstinline[style=mystyle]{<|im_end|>} denote the beginning and end of the dialogue content. During inference, the data format is: ``\lstinline[style=mystyle]{<|im_start|>user<S><D><C><P><|im_end|><|im_start|>assistant}", with the prompt set as: ``Please transcribe the stuttered speech into fluent text based on the provided multimodal information''.

Consistent with standard practices, the loss was computed only on the \lstinline[style=mystyle]{<T>}. The AdamW optimizer was configured with a maximum learning rate of 0.0005 and no weight decay. Training was limited to 25,000 steps. The LoRA parameters were set as follows: rank $r$=8, $alpha$=16, dropout rate $0.1$, and untrained bias terms. The LLM generated outputs autoregressively with a beam width of 2, applying a $repetition\_penalty$ of 1.5 and a $no\_repeat\_ngram\_size$ of 3 to mitigate text repetition. The focal loss balancing factors were assigned as $\alpha=[0.3, 0.3, 0.2, 0.1, 0.1]$, while $\delta=0.3$, $\beta=0.3$, and $\mu=0.1$.

\subsection{Dataset Description}

This study evaluates the proposed methodology using the AS-70 dataset\cite{as-70}, which comprises audio recordings from 70 adults who stutter (24 females, 46 males) and native Mandarin speakers. The dataset includes approximately 50 hours of speech with over 38,000 stuttering events and more than 400,000 transcribed characters, offering both character-level transcriptions and stuttering labels crucial for stuttering recognition and event detection. Recordings were segmented into dialogues and recitations, with each participant contributing one hour of speech. The annotation guidelines categorize stuttering into five types:

\begin{itemize}[label=$\bullet$] 
\item \textbf{\lbrack\rbrack}: Word/Phrase Repetition (complete repetitions).
\item \textbf{/b}: Block (speech interruptions).
\item \textbf{/p}: Prolongation (extended sounds).
\item \textbf{/r}: Sound Repetition (repeated phonemes not forming full characters).
\item \textbf{/i}: Interjections (stuttering-related fillers like ``uh",``um", ``er", excluding natural fillers).
\end{itemize}

\subsection{Results and Discussion}

\subsubsection{ASR Performance}


\autoref{tab:asr} presents the CER results of our LLM-driven ASR-SED framework on the AS-70 test set. CER was evaluated across stuttering severity levels (Mild, Moderate, Severe) and scenarios (Conversation, Command), with comparisons to prior work. We preprocessed the data labels by removing stuttering event markers and disfluent characters. Our experiments analyzed the impact of multimodal embeddings on performance. Results showed that relying solely on speech embeddings degraded performance due to LLM hallucinations caused by stuttering. Incorporating CTC hypothesis embeddings with multimodal speech features suppressed hallucinations and improved robustness. Adding sentence-level stuttering embeddings enabled the LLM to better distinguish stuttered vs. fluent speech, yielding more accurate transcriptions. Compared to the AS-70 baseline, our model achieved a CER of 5.45\% (37.71\% relative reduction).

\subsubsection{SED Performance}

\setlength{\tabcolsep}{3.5pt}
\setlength{\extrarowheight}{1pt}
\begin{table}[htb]
    \caption{Comparison of F1-score (\%) between the method we proposed and existing methods for stuttering event detection}
    \centering
    \begin{tabular}{lcccccc}
        \toprule
        \textbf{Model} & \textbf{/p} & \textbf{/b} & \textbf{/r} & \textbf{[]} & \textbf{/i} & \textbf{Avg} \\
        \midrule
        \multicolumn{7}{l}{\textbf{\textit{Previous works}}} \\
        \midrule
        \textbf{StutterNet}\cite{stutter-net-sheikh2023advancing} & 61.07 & 33.33 & 47.81 & 50.12 & 58.82 & 50.23 \\
        \textbf{Conformer}\cite{gulati2020conformer} & 66.77 & 30.94 & 46.84 & 65.49 & 73.1 & 56.63 \\
        \textbf{ConvLSTM}\cite{convlstm} & 33.30 & 18.22 & 30.19 & 64.02 & 46.70 & 38.49 \\
        \textbf{Wav2vec2.0}\cite{bayerl2022detecting} & 70.48 & 42.51 & 65.76 & 78.48 & 83.80 & 68.21 \\
        \midrule
        \multicolumn{7}{l}{\textbf{\textit{Ours}}} \\
        \midrule
        \textbf{SenseVoice} & 67.07 & 34.56 & 57.7 & 91.85 & 87.69 & 67.77 \\
        \hspace{0.5em}\textbf{+CL} & 71.31 & 38.99 & 64.20 & 93.10 & 87.72 & 71.06 \\
        \hspace{1.5em}\textbf{+Focal Loss} & \textbf{72.79} & \textbf{47.37} & \textbf{66.93} & \textbf{93.29} & \textbf{87.76} & \textbf{73.63} \\
        \bottomrule
    \end{tabular}

    \raggedright 
    \vspace{0.1em} 
    \caption*{
      \makebox[\textwidth][l]{\scriptsize\textbf{CL} denotes multi-label supervised contrastive learning.}
    }
    \vspace{-1em} 
    \label{tab:sed}
\end{table}

\autoref{tab:sed} presents the F1-score results of our proposed method for different stuttering event types on the SED test set of the AS-70 dataset, compared with previous studies. Due to the heterogeneity and overlap inherent in stuttered speech, coupled with limited data and imbalanced class distributions, the SenseVoice model (which utilizes the classifier architecture and binary CE loss as described in \autoref{subsec: sed_branch}
) performs poorly, as do earlier works. However, by introducing two key improvements—multi-label supervised contrastive learning and focal loss—the model's ability to discriminate stuttered speech acoustic features is enhanced, and effectively mitigates the long-tailed distribution in stuttering event categories. Our model significantly outperforms others in recognizing Word/Phrase Repetition ([]) events and shows substantial improvements in detecting the more challenging Block (/b) events, validating the effectiveness of our approach. Ultimately, the model’s average F1 score increases to 73.63\%, representing a relative improvement of 46.58\%.

\subsubsection{Effectiveness of the LLM-Driven ASR-SED Multi-Task Framework}

Our experiments validated the effectiveness of the LLM-driven ASR-SED framework, achieving a CER of 5.45\% (37.71\% relative reduction) and an average SED F1-score of 73.63\% (46.58\% relative gain) on the AS-70 dataset. By dynamically integrating CTC hypothesis embeddings and stuttering embeddings, the framework suppressed repetitive hallucinations caused by stuttered speech while enhancing cross-modal alignment. The hybrid loss combining multi-label contrastive learning and focal loss effectively addressed class imbalance, improving detection of challenging events like Block (/b) and Word Repetition. Importantly, the model maintained robust performance across stuttering severities (Mild/Severe) and scenarios (Conversation/Command), demonstrating its adaptability to real-world applications such as speech rehabilitation. These results confirm that joint optimization of ASR and SED tasks through LLM-driven interaction enables accurate and unified disfluency processing, providing a foundation for extending similar architectures to multilingual or low-resource speech disorder scenarios.

\section{Conclusion}

This paper presents an LLM-driven ASR-SED multi-task learning framework that achieves dual breakthroughs in recognition and detection performance in Mandarin stuttered speech scenarios through dynamic interaction mechanisms and hybrid optimization strategies. Experiments demonstrate that the framework, by integrating CTC hypothesis embeddings, contrastive learning, and focal loss, significantly mitigates the repetitive hallucination issue in stuttered speech and improves the detection of low-frequency event categories. Future work will further explore the potential of LLMs in addressing stuttering and drive the application of this technology in practical rehabilitation settings.

\section{Acknowledgements}
This work is supported by National Engineering Research Center of Multi-dimensional Identification and Trusted Authentication Technology (No.IDNERC202403).

\ifinterspeechfinal
\else
\fi

\bibliographystyle{IEEEtran}
\bibliography{mybib}

\begin{thebibliography}{10}
\providecommand{\url}[1]{#1}
\csname url@samestyle\endcsname
\providecommand{\newblock}{\relax}
\providecommand{\bibinfo}[2]{#2}
\providecommand{\BIBentrySTDinterwordspacing}{\spaceskip=0pt\relax}
\providecommand{\BIBentryALTinterwordstretchfactor}{4}
\providecommand{\BIBentryALTinterwordspacing}{\spaceskip=\fontdimen2\font plus
\BIBentryALTinterwordstretchfactor\fontdimen3\font minus
  \fontdimen4\font\relax}
\providecommand{\BIBforeignlanguage}[2]{{%
\expandafter\ifx\csname l@#1\endcsname\relax
\typeout{** WARNING: IEEEtran.bst: No hyphenation pattern has been}%
\typeout{** loaded for the language `#1'. Using the pattern for}%
\typeout{** the default language instead.}%
\else
\language=\csname l@#1\endcsname
\fi
#2}}
\providecommand{\BIBdecl}{\relax}
\BIBdecl

\bibitem{wu2023world}
S.~Wu, ``“the world is designed for fluent people”: Benefits and challenges
  of videoconferencing technologies for people who stutter,'' in
  \emph{Proceedings of the 2023 CHI Conference on Human Factors in Computing
  Systems}, 2023, pp. 1--17.

\bibitem{coalson2022microaggression}
G.~A. Coalson, A.~Crawford, S.~B. Treleaven, C.~T. Byrd, L.~Davis, L.~Dang,
  J.~Edgerly, and A.~Turk, ``Microaggression and the adult stuttering
  experience,'' \emph{Journal of Communication Disorders}, vol.~95, p. 106180,
  2022.

\bibitem{2-huang2024enhanced}
S.~Huang, D.~Zhang, J.~Deng, and R.~Zheng, ``Enhanced asr for stuttering
  speech: Combining adversarial and signal-based data augmentation,'' in
  \emph{2024 IEEE Spoken Language Technology Workshop (SLT)}.\hskip 1em plus
  0.5em minus 0.4em\relax IEEE, 2024, pp. 393--400.

\bibitem{1-huang2024fosafer}
S.~Huang, D.~Zhang, Y.~Wang, J.~Deng, and R.~Zheng, ``The fosafer system for
  the chime-8 mmcsg challenge,'' in \emph{CHiME Workshop on Speech Processing
  in Everyday Environments}, 2024.

\bibitem{as-70}
R.~Gong, H.~Xue, L.~Wang, X.~Xu, Q.~Li, L.~Xie, H.~Bu, S.~Wu, J.~Zhou, Y.~Qin
  \emph{et~al.}, ``As-70: A mandarin stuttered speech dataset for automatic
  speech recognition and stuttering event detection,'' \emph{arXiv preprint
  arXiv:2406.07256}, 2024.

\bibitem{xue2024findings}
H.~Xue, R.~Gong, M.~Shao, X.~Xu, L.~Wang, L.~Xie, H.~Bu, J.~Zhou, Y.~Qin, J.~Du
  \emph{et~al.}, ``Findings of the 2024 mandarin stuttering event detection and
  automatic speech recognition challenge,'' in \emph{2024 IEEE Spoken Language
  Technology Workshop (SLT)}.\hskip 1em plus 0.5em minus 0.4em\relax IEEE,
  2024, pp. 385--392.

\bibitem{21-zayats2016disfluency}
V.~Zayats, M.~Ostendorf, and H.~Hajishirzi, ``Disfluency detection using a
  bidirectional lstm,'' \emph{arXiv preprint arXiv:1604.03209}, 2016.

\bibitem{22-shonibare2022enhancing}
O.~Shonibare, X.~Tong, and V.~Ravichandran, ``Enhancing asr for stuttered
  speech with limited data using detect and pass,'' \emph{arXiv preprint
  arXiv:2202.05396}, 2022.

\bibitem{23-lea2023user}
C.~Lea, Z.~Huang, J.~Narain, L.~Tooley, D.~Yee, D.~T. Tran, P.~Georgiou, J.~P.
  Bigham, and L.~Findlater, ``From user perceptions to technical improvement:
  Enabling people who stutter to better use speech recognition,'' in
  \emph{Proceedings of the 2023 CHI Conference on Human Factors in Computing
  Systems}, 2023, pp. 1--16.

\bibitem{24-mitra2021analysis}
V.~Mitra, Z.~Huang, C.~Lea, L.~Tooley, S.~Wu, D.~Botten, A.~Palekar,
  S.~Thelapurath, P.~Georgiou, S.~Kajarekar \emph{et~al.}, ``Analysis and
  tuning of a voice assistant system for dysfluent speech,'' \emph{arXiv
  preprint arXiv:2106.11759}, 2021.

\bibitem{25-zhang2022stutter}
X.~Zhang, I.~Vall{\'e}s-P{\'e}rez, A.~Stolcke, C.~Yu, J.~Droppo, O.~Shonibare,
  R.~Barra-Chicote, and V.~Ravichandran, ``Stutter-tts: Controlled synthesis
  and improved recognition of stuttered speech,'' \emph{arXiv preprint
  arXiv:2211.09731}, 2022.

\bibitem{sheikh2021stutternet}
S.~A. Sheikh, M.~Sahidullah, F.~Hirsch, and S.~Ouni, ``Stutternet: Stuttering
  detection using time delay neural network,'' in \emph{2021 29th European
  Signal Processing Conference (EUSIPCO)}.\hskip 1em plus 0.5em minus
  0.4em\relax IEEE, 2021, pp. 426--430.

\bibitem{al2022stuttering}
A.-K. Al-Banna, E.~Edirisinghe, and H.~Fang, ``Stuttering detection using
  atrous convolutional neural networks,'' in \emph{2022 13th International
  Conference on Information and Communication Systems (ICICS)}.\hskip 1em plus
  0.5em minus 0.4em\relax IEEE, 2022, pp. 252--256.

\bibitem{kourkounakis2020fluentnet}
T.~Kourkounakis, A.~Hajavi, and A.~Etemad, ``Fluentnet: end-to-end detection of
  speech disfluency with deep learning,'' \emph{arXiv preprint
  arXiv:2009.11394}, 2020.

\bibitem{liu2024end}
X.~Liu, C.~Xu, Y.~Yang, L.~Wang, and N.~Yan, ``An end-to-end stuttering
  detection method based on conformer and bilstm,'' \emph{arXiv preprint
  arXiv:2411.09479}, 2024.

\bibitem{cl2-2he2020momentum}
K.~He, H.~Fan, Y.~Wu, S.~Xie, and R.~Girshick, ``Momentum contrast for
  unsupervised visual representation learning,'' in \emph{Proceedings of the
  IEEE/CVF conference on computer vision and pattern recognition}, 2020, pp.
  9729--9738.

\bibitem{cl3-zhang2021cola}
C.~Zhang, M.~Cao, D.~Yang, J.~Chen, and Y.~Zou, ``Cola: Weakly-supervised
  temporal action localization with snippet contrastive learning,'' in
  \emph{Proceedings of the IEEE/CVF conference on computer vision and pattern
  recognition}, 2021, pp. 16\,010--16\,019.

\bibitem{mu2024mmger}
B.~Mu, Y.~Li, Q.~Shao, K.~Wei, X.~Wan, N.~Zheng, H.~Zhou, and L.~Xie, ``Mmger:
  Multi-modal and multi-granularity generative error correction with llm for
  joint accent and speech recognition,'' \emph{arXiv preprint
  arXiv:2405.03152}, 2024.

\bibitem{ma2025speech}
Z.~Ma, G.~Yang, Y.~Yang, Z.~Gao, J.~Wang, Z.~Du, F.~Yu, Q.~Chen, S.~Zheng,
  S.~Zhang \emph{et~al.}, ``Speech recognition meets large language model:
  Benchmarking, models, and exploration,'' in \emph{Proceedings of the AAAI
  Conference on Artificial Intelligence}, vol.~39, no.~23, 2025, pp.
  24\,840--24\,848.

\bibitem{li2024transcription}
Y.~Li, X.~Wang, S.~Cao, Y.~Zhang, L.~Ma, and L.~Xie, ``A transcription
  prompt-based efficient audio large language model for robust speech
  recognition,'' \emph{arXiv preprint arXiv:2408.09491}, 2024.

\bibitem{du2023lauragpt}
Z.~Du, J.~Wang, Q.~Chen, Y.~Chu, Z.~Gao, Z.~Li, K.~Hu, X.~Zhou, J.~Xu, Z.~Ma
  \emph{et~al.}, ``Lauragpt: Listen, attend, understand, and regenerate audio
  with gpt,'' \emph{arXiv preprint arXiv:2310.04673}, 2023.

\bibitem{zhang2023speechgpt}
D.~Zhang, S.~Li, X.~Zhang, J.~Zhan, P.~Wang, Y.~Zhou, and X.~Qiu, ``Speechgpt:
  Empowering large language models with intrinsic cross-modal conversational
  abilities,'' \emph{arXiv preprint arXiv:2305.11000}, 2023.

\bibitem{chu2023qwen-audio}
Y.~Chu, J.~Xu, X.~Zhou, Q.~Yang, S.~Zhang, Z.~Yan, C.~Zhou, and J.~Zhou,
  ``Qwen-audio: Advancing universal audio understanding via unified large-scale
  audio-language models,'' \emph{arXiv preprint arXiv:2311.07919}, 2023.

\bibitem{tang2023salmonn}
C.~Tang, W.~Yu, G.~Sun, X.~Chen, T.~Tan, W.~Li, L.~Lu, Z.~Ma, and C.~Zhang,
  ``Salmonn: Towards generic hearing abilities for large language models,''
  \emph{arXiv preprint arXiv:2310.13289}, 2023.

\bibitem{hu2024large}
Y.~Hu, C.~Chen, C.-H.~H. Yang, R.~Li, C.~Zhang, P.-Y. Chen, and E.~Chng,
  ``Large language models are efficient learners of noise-robust speech
  recognition,'' \emph{arXiv preprint arXiv:2401.10446}, 2024.

\bibitem{ctc-graves2006connectionist}
A.~Graves, S.~Fern{\'a}ndez, F.~Gomez, and J.~Schmidhuber, ``Connectionist
  temporal classification: labelling unsegmented sequence data with recurrent
  neural networks,'' in \emph{Proceedings of the 23rd international conference
  on Machine learning}, 2006, pp. 369--376.

\bibitem{focal-ross2017focal}
T.-Y. Ross and G.~Doll{\'a}r, ``Focal loss for dense object detection,'' in
  \emph{proceedings of the IEEE conference on computer vision and pattern
  recognition}, 2017, pp. 2980--2988.

\bibitem{hu2021lora}
E.~J. Hu, Y.~Shen, P.~Wallis, Z.~Allen-Zhu, Y.~Li, S.~Wang, L.~Wang, and
  W.~Chen, ``Lora: Low-rank adaptation of large language models,'' \emph{arXiv
  preprint arXiv:2106.09685}, 2021.

\bibitem{4-an2024funaudiollm}
K.~An, Q.~Chen, C.~Deng, Z.~Du, C.~Gao, Z.~Gao, Y.~Gu, T.~He, H.~Hu, K.~Hu
  \emph{et~al.}, ``Funaudiollm: Voice understanding and generation foundation
  models for natural interaction between humans and llms,'' \emph{arXiv
  preprint arXiv:2407.04051}, 2024.

\bibitem{stutter-net-sheikh2023advancing}
S.~A. Sheikh, M.~Sahidullah, F.~Hirsch, and S.~Ouni, ``Advancing stuttering
  detection via data augmentation, class-balanced loss and multi-contextual
  deep learning,'' \emph{IEEE Journal of Biomedical and Health Informatics},
  vol.~27, no.~5, pp. 2553--2564, 2023.

\bibitem{gulati2020conformer}
A.~Gulati, J.~Qin, C.-C. Chiu, N.~Parmar, Y.~Zhang, J.~Yu, W.~Han, S.~Wang,
  Z.~Zhang, Y.~Wu \emph{et~al.}, ``Conformer: Convolution-augmented transformer
  for speech recognition,'' \emph{arXiv preprint arXiv:2005.08100}, 2020.

\bibitem{convlstm}
C.~Lea, V.~Mitra, A.~Joshi, S.~Kajarekar, and J.~P. Bigham, ``Sep-28k: A
  dataset for stuttering event detection from podcasts with people who
  stutter,'' in \emph{ICASSP 2021-2021 IEEE International Conference on
  Acoustics, Speech and Signal Processing (ICASSP)}.\hskip 1em plus 0.5em minus
  0.4em\relax IEEE, 2021, pp. 6798--6802.

\bibitem{bayerl2022detecting}
S.~P. Bayerl, D.~Wagner, E.~N{\"o}th, and K.~Riedhammer, ``Detecting
  dysfluencies in stuttering therapy using wav2vec 2.0,'' \emph{arXiv preprint
  arXiv:2204.03417}, 2022.

\end{thebibliography}

\end{document}